\documentclass[journal=jacsat,manuscript=article]{achemso}

\usepackage[version=3]{mhchem} 
\usepackage[dvipsnames]{xcolor}
\usepackage[normalem]{ulem}
\usepackage{amsfonts}
\usepackage{microtype} 

\author{Kai Töpfer}
\affiliation[]
{Department of Biology, Chemistry and
Pharmacy, Freie Universität Berlin, 14195 Berlin, Germany.}
\alsoaffiliation{These authors contributed equally to this work.}
\author{Gianmarco Lazzeri}
\affiliation{Institute of Computer Science, Goethe University Frankfurt, Frankfurt am Main, Germany.}
\alsoaffiliation[FIAS]
{Frankfurt Institute for Advanced Studies, Frankfurt am Main, Germany.}
\alsoaffiliation{These authors contributed equally to this work.}
\author{Vittoria Ossanna}
\affiliation[Second University]{Department of Cellular, Computational and Integrative Biology, University of Trento, Trento, Italy.}
\alsoaffiliation[FIAS]
{Frankfurt Institute for Advanced Studies, Frankfurt am Main, Germany.}
\author{Florian Renner}
\affiliation[]
{Department of Biology, Chemistry and
Pharmacy, Freie Universität Berlin, 14195 Berlin, Germany.}
\author{Gianluca Lattanzi}
\affiliation[]{Deparment of Physics, University of Trento, Trento, Italy.}
\author{Roberto Covino}
\affiliation[Unknown University]
{Institute of Computer Science, Goethe University Frankfurt, Frankfurt am Main, Germany.}
\alsoaffiliation[FIAS]
{Frankfurt Institute for Advanced Studies, Frankfurt am Main, Germany.}
\author{Bettina G.~Keller}
\affiliation[]
{Department of Biology, Chemistry and
Pharmacy, Freie Universität Berlin, 14195 Berlin, Germany.}
\email{bettina.keller@fu-berlin.de, covino@fias.uni-frankfurt.de}

\title[An \textsf{achemso} demo]
{A Machine-Learned Symbolic Committor for a Chemical Reaction: Retinal Isomerization}

\begin{document}

\begin{tocentry}
\includegraphics[width=8.25cm]{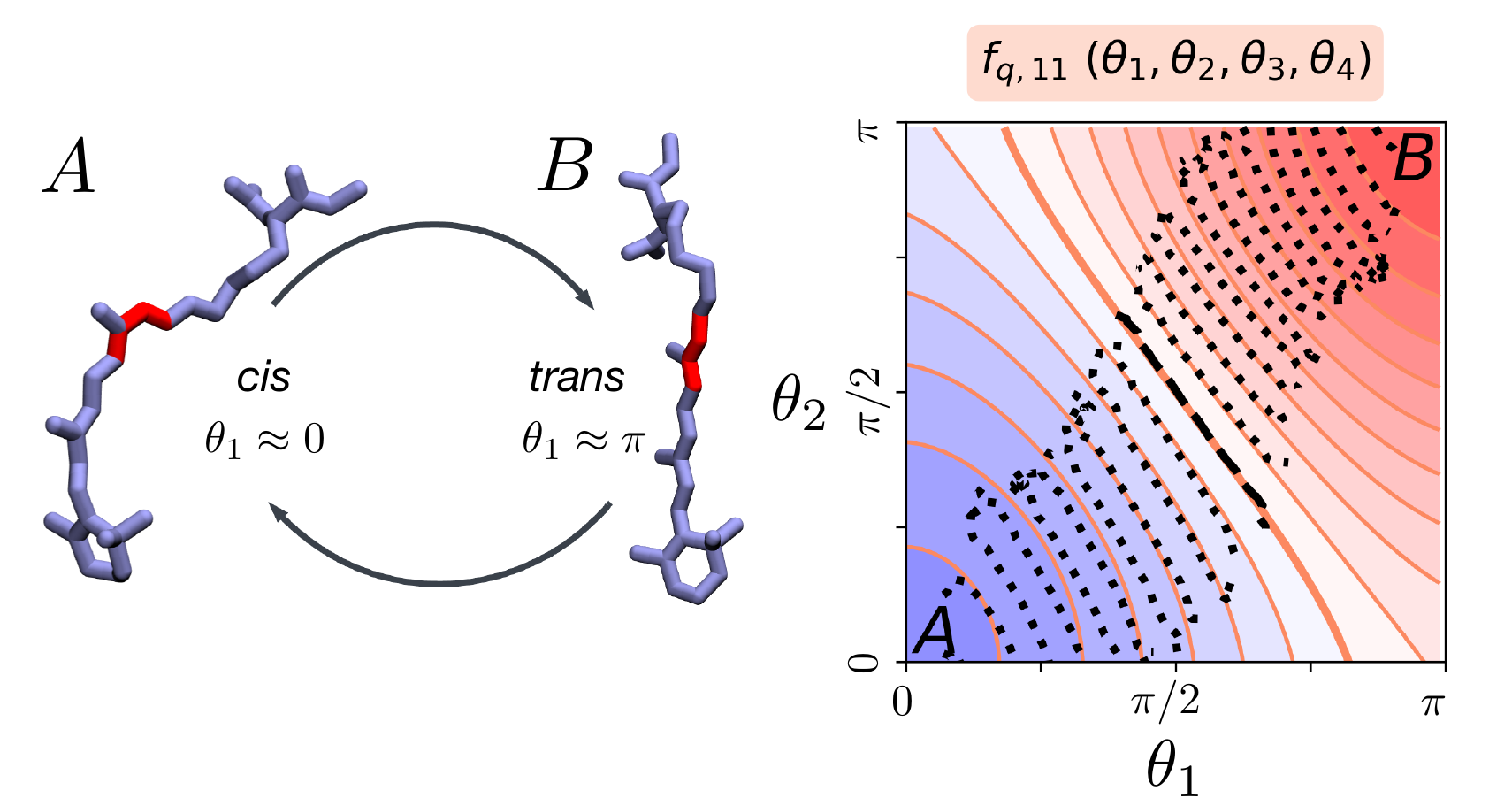}
\end{tocentry}

\begin{abstract}
The thermal \textsl{cis}–\textsl{trans} isomerization around the C$_{13}$=C$_{14}$ double bond of retinal is a prototypical high-barrier reaction whose mechanism hinges on subtle out-of-plane bending motions. We apply Artificial Intelligence for Molecular Mechanism Discovery (AIMMD) to N-retinylidene-lysine in vacuum, learning the committor from unbiased molecular dynamics trajectories generated by two-way shooting. Parametrizing the logit of the committor, rather than the committor itself, allows the neural network to resolve the reaction coordinate across the full transition region, not only at the isocommittor surface $p_B(\mathbf{x}) = 0.5$. Holdback input randomization identifies four proper dihedrals around the reactive bond as the informative coordinates, while the improper dihedrals at C$_{13}$ and C$_{14}$ prove unsuitable because reactant, transition, and product states share the same values. Symbolic regression then distills the network into compact analytical expressions and shows that a nonlinear coupling of all four dihedrals is required to reproduce the S-shaped, stepwise pathway seen in the transition path ensemble. This S-shape is absent from the minimum-free-energy path: it arises from the non-equilibrium dynamics of the short ($\sim 0.13$\,ps) transition events combined with the mass asymmetry between heavy-atom and hydrogen-bearing dihedrals. An interpretable, machine-learned committor thus exposes dynamical features of the mechanism to which the free-energy surface is blind. The workflow requires no prior assumptions about the reaction coordinate and extends naturally to other isomerizations and to chemical reactions more broadly.
\end{abstract}

\newpage
\section{Introduction}

A rare event \cite{peters2017reaction, frenkel2023understanding} is a transition across an energy barrier that occurs infrequently  because the barrier is much higher than the average thermal fluctuations.
Chemical reactions are a prototypical but also a particularly challenging class of rare events 
\cite{bolhuis2002transition, peters2016reaction}. 
To rationalize their mechanism and to calculate reaction rates, one introduces reaction coordinates, i.e.~functions of the conformational space that monotonically scale between the two energy basins in either side of the barrier.
While it is usually possible to define a naive reaction coordinate by simply interpolating between reactant and product state, such a reaction coordinate poorly captures the mechanism and often yields highly inaccurate rates.
Finding optimal reaction coordinates is therefore a long-standing but still very active field of research \cite{chen2023discovering, gkeka2020machine, beyerle2022quantifying}.
Minimum energy paths \cite{maragliano2006string,perez2019adaptive,diaz2012path}, traditionally used as models for the optimal reaction coordinate, identify the transition state (TS) as a saddle point in the PES but neglect entropic contributions and dynamical effects.
Alternatively, the committor function\cite{berezhkovskii2019committors,covino2019molecular,kang2024computing} characterizes the actual dynamical flow of reactive trajectories.
This offers a complementary perspective, in which the TS is defined as the isocommittor surface at which trajectories are equally likely to reach reactant and product. 
The committor thus yields the mechanism at finite temperature and accurate rate predictions in complex, high-dimensional systems.
Committor functions can be parametrized from molecular dynamics (MD) trajectories, but this is notoriously difficult \cite{bolhuis2002transition, elber2017calculating, mitchell2024committor}, because the committor is a non-linear function of the high-dimensional molecular state space.  
Neural networks have been transformative in this context\cite{ma2005automatic, contreras2026learning, kang2026committors}, as they can efficiently represent high-dimensional, non-linear functions.
Some of us recently developed ``Artificial Intelligence for Molecular Mechanism Discovery'' (AIMMD),\cite{jung2023machine,lazzeri2023aimmd} a neural network that learns the committor of a rare event from short dynamically unbiased trajectories, which we generate using molecular simulations. A low-dimensional approximation of the committor is obtained by symbolic regression, a genetic algorithm that searches both functional form and parameters to best fit a data set. This provides a human-interpretable reaction mechanism expressed in terms of a few internal degrees of freedom, making complex reaction dynamics easier to understand. 

\begin{figure}
    \centering
    \includegraphics[width=1.0\linewidth]{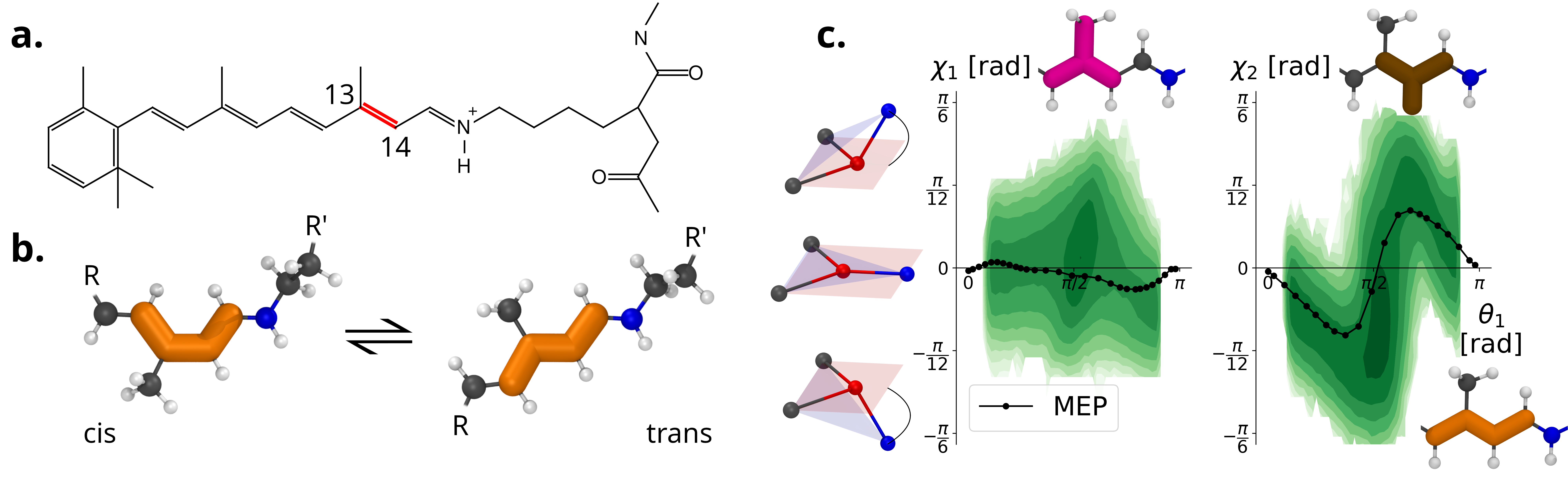}
    \caption{
    {\bf (a.)} Lewis structure of retinal covalently linked to a lysine via a protonated Schiff base (N-retinylidene-lysine). The reaction center around the C$_{13}$=C$_{14}$ double-bond is shown in red with atom labeling. 
    {\bf (b.)} \textsl{Cis}-\textsl{trans} isomerization around the C$_{13}$=C$_{14}$ double-bond, with $\theta_1$ as naive reaction coordinate marked in orange.
    {\bf (c.)} Minimum energy path (black) and
    logarithmic transition path ensemble (green) of reactive trajectories projected into the space of the naive reaction coordinate $\theta_1$ and the improper dihedral angles, $\chi_1$ and $\chi_2$, centered at C$_{13}$ and C$_{14}$, respectively.
    }
    \label{fig:problemStatement}
\end{figure}

AIMMD has been successfully used to model a variety of diffusive molecular self-organization processes.\cite{jung2023machine,lazzeri2023aimmd,jackel2025free,post2025ai,horvath2025stim1}
However, it remains unclear whether learning the committor for chemical reactions is possible or practical, for two reasons:
\textbf{1:} For the high free-energy barriers that characterize chemical reactions, the committor is essentially a step function that switches between 0 (reactant) and 1 (product) with only a narrow region around the TS yielding intermediate values. 
A model that accurately predicts only the isocommittor surface $p_B(\mathbf{x})=0.5$ will yield highly accurate predictions for most data points in a test data set. 
However, information on the reaction mechanism hinges on the how accurately the few data points in the intermediate committor region are represented.
A model that accurately predicts only the isocommittor surface $p_B(\mathbf{x})=0.5$ misses crucial mechanistic information in the vast majority of the reactive region, as it does not capture twists and turns of the reactive channels. Learning away from the transition state becomes progressively more demanding, since ordering configurations along the reaction coordinate is increasingly difficult near the reactant and product basins, where the committor gradient vanishes.\cite{breebaart2025understanding} 

\textbf{2:} In transitions with high energy barriers the transition event is very short. %
As a result, the assumption of instantaneous momentum relaxation may not hold during barrier crossing, and accurate description of the transition-path ensemble might require inertial dynamics.
Consequently, a committor function that depends only on positions and neglects momenta might provide an incomplete description of the reaction.
As a challenging test case for AIMMD, we model the thermal \textsl{cis}–\textsl{trans} isomerization around the C$_{13}$=C$_{14}$ double bond in retinal.
Specifically,  we consider the protonated retinal Schiff base covalently linked to a lysine residue, i.e.~N-retinylidene-lysine, in vacuum (Fig.~\ref{fig:problemStatement}a and Fig.~\ref{fig:reaction_center}).
It has a free-energy barrier\cite{ghysbrecht2025accuracy} of $\sim 100$\,kJ/mol (in our force-field model) and likely exhibits the two challenges discussed above. 
In addition is has a complicated reaction coordinate.
The intuitive reaction coordinate is the dihedral angle $\theta_1$ (Fig.~\ref{fig:problemStatement}b) which cleanly separates reactant, transition, and product states but fails to accurately parametrize the probability flux across the TS. \cite{ghysbrecht2024thermal, ghysbrecht2025accuracy}
This is because near the TS, the reaction proceeds via out-of-plane bending motions of C$_{13}$ and C$_{14}$ (Fig.~\ref{fig:problemStatement}c), which have to be included in order to obtain accurate rate estimates \cite{ghysbrecht2025accuracy}. 
These out-of-plane motions add two additional challenges. 
\textbf{3:} The amplitudes of these motions are small and therefore difficult to detect in trajectory data. 
Accurately capturing them requires a model with sufficient resolution.
\textbf{4:} The out-of-plane bending is a flip-flop motion (Fig.~\ref{fig:problemStatement}c), where the molecule forms a pyramidal distortion in one direction and, upon crossing the TS, reverses the distortion to the opposite side \cite{ghysbrecht2024thermal}.
The improper dihedrals $\chi_1$ and $\chi_2$ describe this motion.
However, while the committor increases monotonically along reactive trajectories and thus provides a consistent measure of reaction progress, $\chi_1$ and $\chi_2$ do not.
Instead, they can take similar values in reactant, transition, and product states.
Consequently, they provide little information about reaction progress and are therefore not suitable for parametrizing the committor.
We present an in-depth analysis of how a neural-network model of the committor responds to these challenges.
In this work, we show how a neural-network model, employed within an active learning framework, can both efficiently sample rare-event trajectories and yield a versatile, interpretable model of the committor that discloses the reaction mechanism of retinal isomerization beyond the transition state.



%
%
\section{Theory}

Consider a system with $N$ atoms, where $\mathbf{x} \in \mathbb{R}^{3N}$ is a state in the $3N$-dimensional positional space. 
We define two regions $A, B \subset \mathbb{R}^{3N}$ with $A$ representing the reactant state and $B$ representing the product state.
(See Eq.~\ref{eq:state_definition_function} for the definitions in retinal.)
The system evolves according to a dynamical process for which the first hitting times $\tau_A$ and $\tau_B$ of 
of the regions $A$ and $B$ are well defined for all states outside these regions ($\mathbf{x} \not\in A \cup B$) (effectively integrating out the velocities).
The committor probability to the product state $B$ for a process $\mathbf{x}_t$ starting in $\mathbf{x}_0$ at time $t=0$ is
\begin{align}
    p_{B}(\mathbf{x}) = P(\tau_B < \tau_A | \mathbf{x}_0 = \mathbf{x}) \, .
\label{eq:committor}    
\end{align}
For any conformation within the reactant state $A$, the first hitting time satisfies $\tau_A\stackrel{\mathrm{def}}{=} 0$ and hence $p_B(\mathbf{x}) = 0$.
Likewise, for any  within the product state $B$, $\tau_B\stackrel{\mathrm{def}}{=} 0$ and hence $p_B(\mathbf{x}) = 1$.
The complement of $p_B(\mathbf{x})$ is the committor function to the reactant state $p_A(\mathbf{x}) = 1 - p_B(\mathbf{x})$.
Intuitively, Eq.~\ref{eq:committor} represents the probability that a trajectory starting in $\mathbf{x}$ reaches $B$ before $A$.
The committor can be parametrized using the logistic function
\begin{align}
    p_{B}(\mathbf{x}) = \sigma(q) &:= \frac{1}{1+e^{-q(\mathbf{x})}} \cr
    q_B(\mathbf{x}) = \sigma^{-1}(p_B) &:= \ln \dfrac{p_{B}(\mathbf{x} )}{1- p_{B} (\mathbf{x})}\, , 
\label{eq:logit_committor}   
\end{align}
where $q_B(\mathbf{x})$ is the logit of the committor probability to the product state $B$, or in short: the logit committor.
The transformation $\sigma(q_B)$ and its inverse $\sigma^{-1}(p_B)$ ensure that one can readily switch between the committor and the logit committor.

Evolving a single trajectory initiated at $\mathbf{x}$ is often referred as ``(one-way) shooting'',\cite{bolhuis2002transition} and $\mathbf{x}$ as the ``shooting point''. Initializing two trajectories from $\textbf{x}$, time-reversing one of the two due to microscopic time reversibility, and joining them together in a single reactive trajectory constrained to the shooting point is called ``two-way shooting''. A one-way shooting result can be modeled as a Bernoulli trial with the two possible outcomes ``$B$ is reached before $A$'' and ``$A$ is reached before $B$'', occurring with probabilities $p_B(\mathbf{x})$ and $1-p_B(\mathbf{x})$, respectively.
The committor can therefore be inferred from trajectory outcomes using a maximum-likelihood approach.

We represent the logit committor by a parametric model $q_B(\mathbf{x}|\mathbf{w})$ with parameters $\mathbf{w}$, which induces a parametric model of the committor $p_B(\mathbf{x}|\mathbf{w})$ via Eq.~\ref{eq:logit_committor}.
The training data consist of $2k$ trajectory outcomes $\delta_{ij}$ obtained from two-way shooting initialized from shooting points $\mathbf{x}_i$, where both forward and backward trajectories are treated as independent forward trajectories: $j=1,2$.
Since the trajectories are generally initiated from $k$ different configurations, each trial is characterized by its specific committor probability $p_B(\mathbf{x}_i | \mathbf{w})$, and we have two trials per shooting point initialized with opposite starting velocities.  
The negative log-likelihood of the data is therefore not binomial but given by the product of the Bernoulli probabilities of the individual outcomes,
\begin{align}
    l(\mathbf{w} | S) 
    &:=-\ln P(S|\mathbf{w}) \cr
    &= -\ln \prod_{i=1}^{k}\prod_{j=1}^{2} p_B(\mathbf{x}_i|\mathbf{w})^{\delta_{ij}} \cdot (1-p_B(\mathbf{x}_i|\mathbf{w}))^{1-\delta_{ij}} \cr
    &= \sum_{i=1}^{k}\sum_{j=1}^{2} \delta_{ij} \ln (1+e^{-q(\mathbf{x}_i|\mathbf{w})}) + (1-\delta_{ij}) \ln (1+e^{+q(\mathbf{x}_i|\mathbf{w})}) \cr
    &= \sum_{i=1}^{k} r_{i,B} \ln (1+e^{-q(\mathbf{x}_i|\mathbf{w})}) + r_{i,A} \ln (1+e^{+q(\mathbf{x}_i|\mathbf{w})})
\label{eq:loss_function}    
\end{align}
where $\delta_{ij}$ is a binary outcome variable with $\delta_{ij}=1$ if the trajectory reaches $B$ before $A$, and $\delta_{ij}=0$ otherwise. 
We substituted $p_B(\mathbf{x}_i|\mathbf{w})$ with Eq.~\ref{eq:logit_committor} (See supporting info). 
The last line converts the binary trajectory outcome variables $\delta_{i1}$ and $\delta_{i2}$ into 
the shooting result $\mathbf{r}_i=(r_{i,A}, r_{i,B})$ per shooting point, where $\mathbf{r}_i=(2,0)$ if both trajectories end in $A$, $\mathbf{r}_i=(1,1)$ of one trajectory ends in $A$ and the other in $B$, and $\mathbf{r}_i=(0,2)$ if both trajectories end in $B$.
Eq.~\ref{eq:loss_function} defines the loss function $l(\mathbf{w} | S)$ and 
$S = \{(\mathbf{x}_i, \mathbf{r}_i)\}_{i=1}^k$
is the data set consisting of shooting points $\mathbf{x}_i$ and corresponding shooting result $\mathbf{r}_i$.
The data set $S$ is generated by sampling transition paths and rare-event excursion paths using two-way shooting \cite{dellago1998efficient, bolhuis2002transition} from selected shooting points $\mathbf{x}_i$. 
The probability that a sampled path is a transition path depends on the shooting point via \cite{hummer2004transition} $P(\mathrm{TP} | \mathbf{x}) = 2 p_B(\mathbf{x})(1-p_B(\mathbf{x}))$.
This probability has a maximum at $p_B(\mathbf{x})=0.5$. 
By biasing the selection of shooting points according to $p_B(\mathbf{x})$, one can control the sampling rate of TPs \cite{hummer2004transition, peters2006obtaining, lazzeri2023aimmd}. 
%
%
\section{Results and discussion}

\subsection{AIMMD}

\begin{figure}[htb]
    \centering
    \includegraphics[width=8cm]{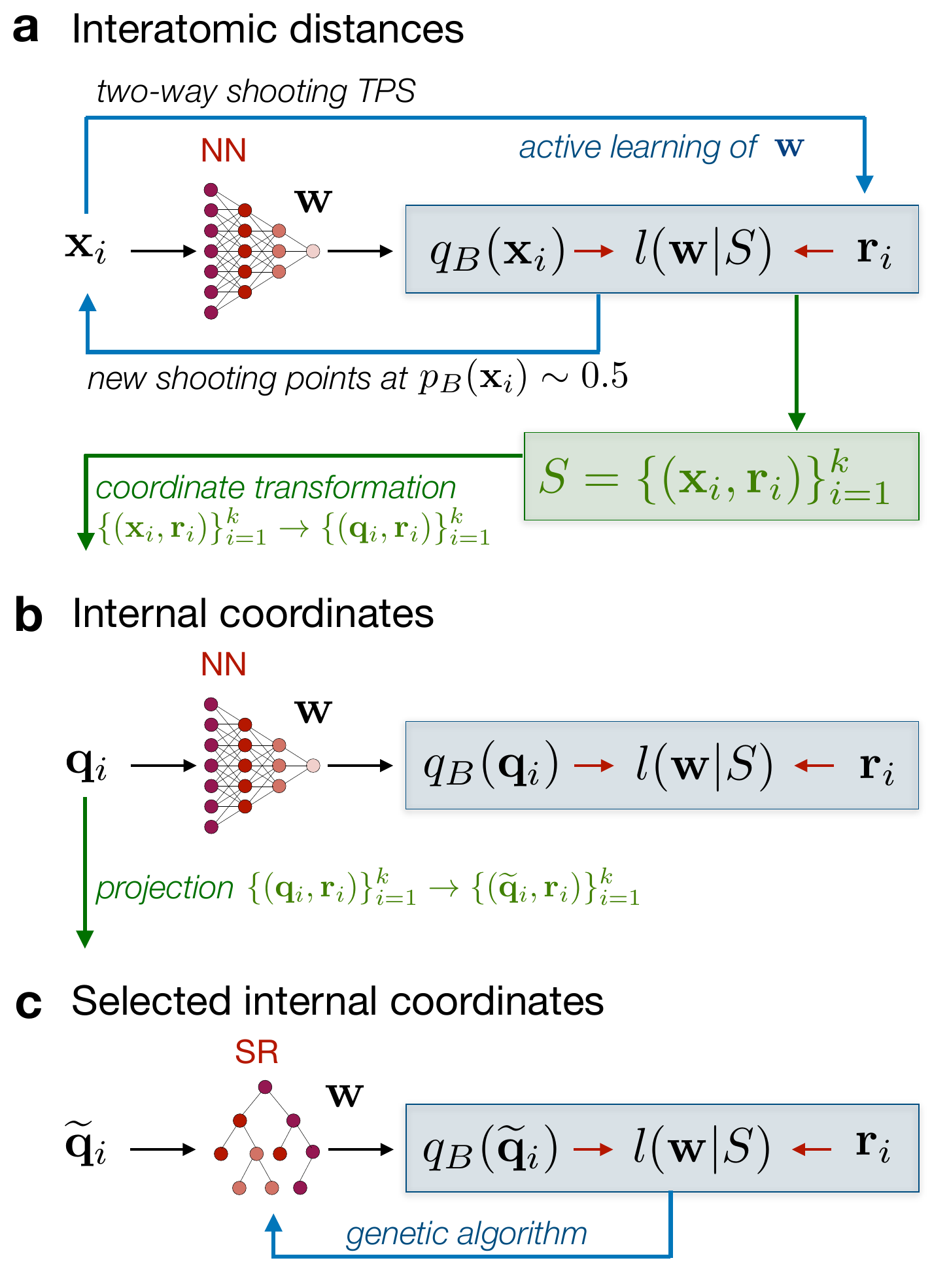}
    \caption{AIMMD. 
    a) Neural-network representation of the logit-committor as a function of Cartesian coordinates $\mathbf{x}$ (represented in terms of pairwise distances) and generation of the path ensemble $\{(\mathbf{x}_i, \mathbf{r}_i)\}_{i=1}^k$;
    b) HIPR analysis using a neural-network representation of the logit-committor as a function of internal coordinates $\mathbf{q}$;
    c) Symbolic regression of the logit-committor as a function of a reduced set of internal coordinates $\widetilde{\mathbf{q}}_i$.
    }
    \label{fig:AIMMD}
\end{figure}

AIMMD is based on a active-learning cycle (Fig.~\ref{fig:AIMMD}.a).
This stage serves two purposes: ($i$) it generates a path ensemble consisting of transition paths and rare-event excursion paths, and ($ii$) it provides a reference neural-network representation of $q_B(\mathbf{x} | \mathbf{w} )$.
The data set $S = \{(\mathbf{x}_i, \mathbf{r}_i)\}_{i=1}^k$ is directly derived from the path ensemble and remains fixed during the subsequent analysis (Fig.~\ref{fig:AIMMD}.b and c). 

\subsubsection{Generating the path ensemble}
The active-learning cycle (Fig.~\ref{fig:AIMMD}.a) starts by training a NN of the logit committor $q_B(\mathbf{x} | \mathbf{w})$, where $\mathbf{w}$ are the NN parameters, from one or several initial transition paths, obtained by, for instance, biased dynamics or interpolation. The NN subsequently guides the sampling of additional paths \cite{jung2023machine} by selecting shooting point configurations $\mathbf{x}_i$ from previously generated reactive trajectories. From each selected configuration, new trajectories are launched using two-way shooting as in transition path sampling (TPS) \cite{falkner2025revisiting,bolhuis2002transition}.
The selection probability is biased toward configurations close to the TS ($p_{B}=0.5$), thereby sampling in high-energy configurations where accurate estimation of the logit-committor is most critical.
In contrast to standard TPS, shooting points can be drawn from all simulations, not only from previously accepted transitions, which accelerates exploration of the reactive region even further \cite{lazzeri2025optimal}. Trajectories initiated from shooting points are terminated as soon as they reach the boundary of either region $A$ or $B$, consistent with the relaxation time of the states, which renders them computationally inexpensive \cite{van2003novel}. Because no biasing forces are applied, the method reproduces the same rare excursions and transition events as an infinitely long unbiased simulation, albeit at a substantially higher rate \cite{lazzeri2023aimmd}. Out of 32,727 generated paths, 9.1\% were transitions paths and constitute the transition path ensemble (TPE). In principle, the transition yield could be further increased; however, this would come at the expense of stronger path correlations and consequently slower exploration of reactive phase space, resulting in a worse committor learning performance especially away from the TS.
The TPE shows the previously observed\cite{ghysbrecht2024thermal, ghysbrecht2025accuracy} zig-zag path in the $(\theta_1, \chi_1)$ and $(\theta_1, \chi_2)$ spaces (Fig.~\ref{fig:problemStatement}.c), confirming that TPS captures a similar reaction mechanism to that obtained with the path-collective-variable approach \cite{ghysbrecht2025accuracy}: torsion around the C$_{13}$=C$_{14}$ double bond with out-of-plane bending at the C$_{13}$ and C$_{14}$ carbon. 
This agreement allows us to use the TPE to further investigate the origin and consequences of out-of-plane bending.

\subsubsection{Reference NN representation of the logit committor}
\begin{figure}[htb]
    \centering
    \includegraphics[width=1\linewidth]{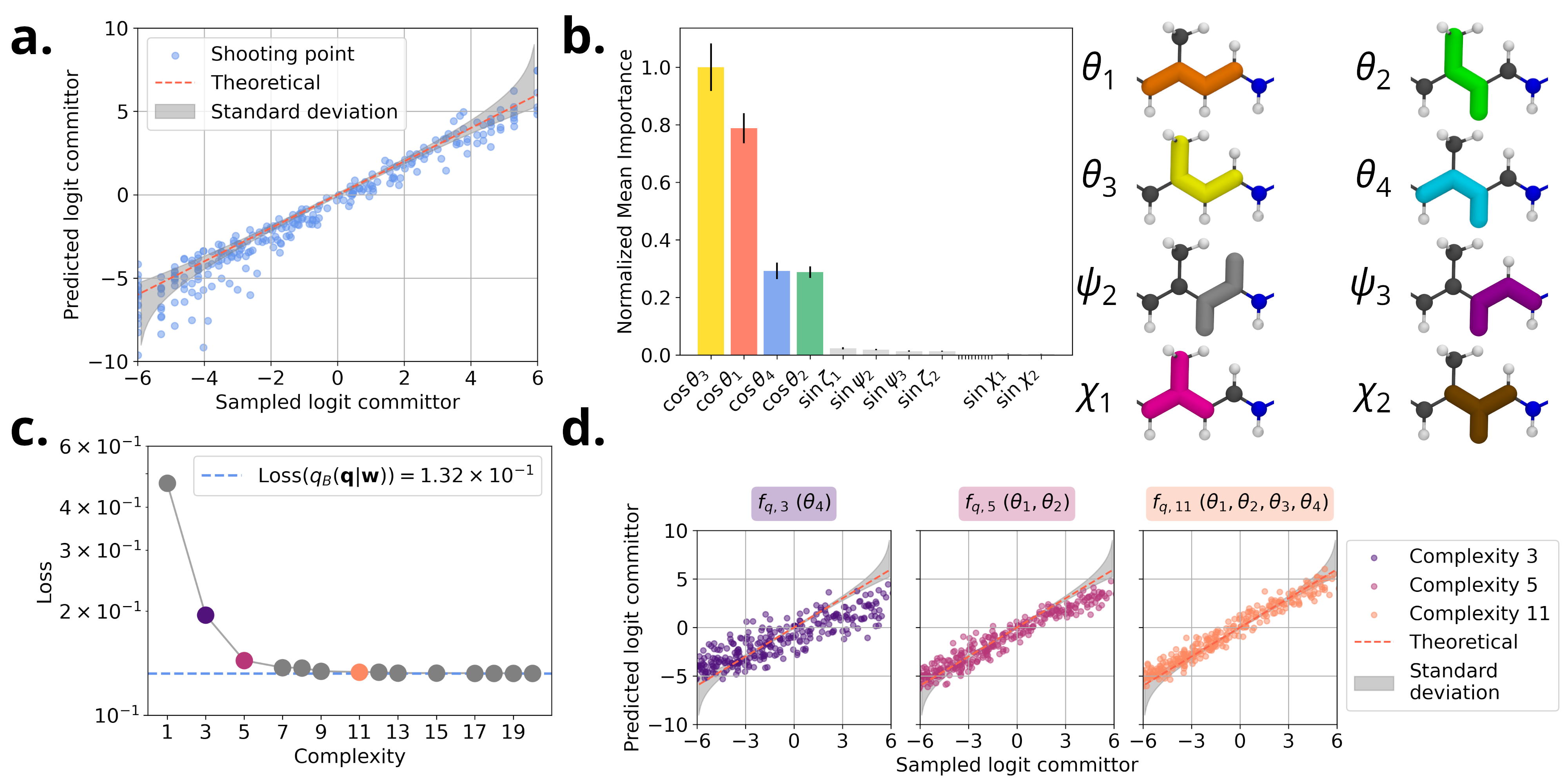}
    \caption{
    {\bf (a.)} Correlation between sampled committor values and NN committor predictions of shooting point conformations (blue dots). 
    {\bf (b.)} Normalized absolute importance of the top 8 features from the HIPR analysis and the most important features of improper angles $\chi_1$ and $\chi_2$. Dihedral angles $\zeta_1$ and $\zeta_2$ includes atoms not included in the atomic sketches, respectively. The black line indicates the standard deviation of the importance from the 10 individual runs.
    {\bf (c.)} Comparison of best loss with increasing of complexity of the committor function predictions from SR.
    {\bf (d.)} Correlation between sampled committor values and committor predictions of shooting point conformers from the best SR functions of complexity (left to right) 3, 5 and 11, respectively.
        }
    \label{fig:NN_representation}
\end{figure}

To ensure that neither $q_B(\mathbf{x} | \mathbf{w} )$ nor $S$ is biased by prior assumptions about the reaction mechanism, the reference NN is trained on agnostic coordinates, namely the interatomic distances of the shooting-point conformations.
Due to the high-free energy barrier in the retinal isomerization reaction, the committor approaches a step function, with values significantly different from 0 and 1 confined to a narrow region around the TS \cite{banushkina2016optimal}. This sharp separation makes accurate learning of the $p_{B}=0.5$ isocommittor surface challenging. At the same time, complex reactive channels involve multiple nonlinear collective motions occurring sequentially \cite{ghysbrecht2024thermal}, which span regions associated with both low and high committor values. Capturing only the TS region is therefore insufficient to describe the full reactive mechanism. To improve numerical stability and representation quality, we expressed the committor $p_{B}$ in terms of the logit variable $q_B$.
In this representation, the reaction coordinate $q({\bf x})$ preserves variations in the input features more uniformly across configuration space ${\bf x}$.
This avoids a limitation of the committor $p_{B}$, whose gradients vanish when approaching states $A$ and $B$, making learning of the committor in these regions less effective. 
Due to the expressivity of $q_B(\mathbf{x})$, the NN that is trained during the active-learning cycle in Fig.~\ref{fig:AIMMD}.a has a realistic chance to represent the reaction coordinate beyond the narrow TS region, provided that the training data are sufficiently diverse and extend away from it. In AIMMD, this goal is achieved by sampling approximately uniformly in the logit of the committor between two set interfaces, $q_{B,\mathrm{min}}=-20$ and $q_{B,\mathrm{max}}=+20$.
Note that this reflects in a comparable number of shooting points being selected between $p_B = 0.001$ and $p_B=0.01$ and between $p_B = 0.01$ and $p_B=0.1$.
To validate this reference model, we estimated the logit committor for 442 starting structures $\mathbf{x}_i$ from an independent AIMMD run using 200 two-way-shooting paths per $\mathbf{x}_i$ and compared the estimates to the NN prediction (Fig.~\ref{fig:NN_representation}.a).
The comparison demonstrates that the model is highly accurate in the interval $q_B(\mathbf{x})\in [-4,4]$ which corresponds to a committor probability interval of $p_B(\mathbf{x}) \in [0.02, 0.98]$.
Estimates and predictions align particularly well in the central region of the plot near the TS.
The slight deviations in  the left and right regions of the plot appear as vertically stacked data points. 
This reflects finite-sampling effects in the committor estimator: while the NN model produces continuous predictions, the estimated committor values become effectively discretized for configurations strongly committed to either state, where only few shooting trajectories reach the opposite state.
The committor distribution of shooting points near the transition state follows the expected Gaussian distribution centered at $0.5$ (Fig.~\ref{fig:halfcmt}), further indicating that the transition state ensemble has been identified accurately.

\subsubsection{Identifying relevant internal degrees of freedom}
Although the reference model accurately approximates the logit committor, it does not directly reveal the reaction center, the internal degrees of freedom that change during the reaction, or the sequence in which these structural changes occur.
To extract the underlying reaction mechanism, we use the data set to train a second neural network (Fig.~\ref{fig:AIMMD}.b), but now represent the input in internal coordinates $\mathbf{q}$, including distances, angles, proper and improper dihedral angles. 

We use the same loss function as in the reference model (Eq.~\ref{eq:loss_function}). 
Thus, the NN in Fig.~\ref{fig:AIMMD}.a and \ref{fig:AIMMD}.b differ only by a coordinate transformation of the input data $\mathbf{x} \rightarrow \mathbf{q}$. 
Fig.~\ref{fig:logit_committor_comparison} shows that the new model $q_B(\mathbf{q} | \mathbf{w})$ reproduces the sampled logit committor even more accurately than the reference model.
To evaluate the relevance of individual internal coordinates for the NN prediction, we then apply the holdback input randomization (HIPR) method \cite{kemp2007approach}, as in Ref.~\citenum{jung2023machine}. 
In this analysis, the values of a specific internal coordinate $q_k$ in the input data set are replaced by uniformly distributed random numbers within an interval $[\min q_k, \max q_k ]$ and the average loss of the modified input data set is calculated. 
Since the modified data set no longer reflects the learned relationship between $\mathbf{q}$ and $q_B(\mathbf{q}| \mathbf{w})$, the loss is expected to increase relative to the loss obtained for the unmodified data. 
The magnitude of this increase indicates the importance of the corresponding coordinate for the NN representation of the logit committor.
To reduce the impact of stochastic fluctuations, the procedure is repeated ten times for each internal coordinate and the loss is evaluated in each cycle.
Crucially, HIPR identifies coordinates that are predictive of the logit committor. 
It does not assess whether a coordinate is geometrically meaningful or intuitive for describing the reaction pathway.
Only four internal degrees of freedom ($\cos\theta_1, \cos\theta_2, \cos\theta_3, \cos\theta_4$) exhibit a noticeable spike in importance in the HIPR analysis, whereas all other internal degrees of freedom have hardly any predictive power for the logit committor (Fig.~\ref{fig:NN_representation}.b). 
The four proper dihedral angles $\theta_1$ to $\theta_4$  (illustrated on the right of Fig.~\ref{fig:NN_representation}.b with matching color codes) correspond to the four possibilities to define the torsion around the C$_{13}$=C$_{14}$ double bond. 
Thus, HIPR correctly identifies the reaction center. 
The fact that neighboring torsion angles (proper dihedrals) contribute little to the model is consistent with previous analyses \cite{ghysbrecht2025accuracy}, which showed that the reaction center is very localized around the C$_{13}$=C$_{14}$-bond.
The improper dihedrals $\chi_1$ and $\chi_2$ at C$_{13}$ and C$_{14}$, respectively, exhibit low predictive relevance for the logit committor. 
This is expected, as the committor is not a monotonic function of $\chi_1$ and $\chi_2$, making them poor coordinates for parametrizing $q_B(\mathbf{x})$. 
In particular, reactant, transition, and product states all share $\chi_1 = \chi_2 = 0$, so this value provides little information about the progress of the reaction.
If the model $q_B(\mathbf{q} \mid \mathbf{w})$ depends only on proper dihedral angles, does it still capture the out-of-plane bending at C$_{13}$ and C$_{14}$? 
In principle, the out-of-plane motion can be represented by a combination of proper dihedrals around the C$_{13}$=C$_{14}$ bond.
The fact that the model relies on all four neighboring proper dihedrals suggests that the reaction coordinate is not trivially parametrized by a single torsion angle. 
However, HIPR does not resolve potential correlations between informative coordinates. 
Even though the high predictive quality of the model (Fig.~\ref{fig:logit_committor_comparison}) indicates that the out-of-plane bending is likely included, we require an explicit representation of the logit committor as a function of internal coordinates to unambiguously identify the relevant degrees of freedom.
Such a representation can be obtained via symbolic regression.


\subsubsection{Explicit representation of the logit committor}

Symbolic regression (SR) is applied to express the logit committor function in a human-interpretable form using the most relevant internal degrees of freedom, determined by HIPR, as input.\cite{makke2024interpretable,jung2023machine}
SR uses a genetic algorithm to both find the best representations in function and parameter space with respect to minimizing the same loss function (Eq.~\ref{eq:loss_function}) as used for the NN models.
The SR results are ordered by complexity (Fig.~\ref{fig:NN_representation}.c) which is measured as the total number of input variables, tunable parameter, binary operators (addition, subtraction, multiplication, division) and unary operators (e.g. exponential or logarithmic function).
The SR algorithm is run 10 times to observe repeating patterns in the results.
With increasing complexity, the estimated formulas by SR include higher number of input variables of which the least complex and best representation with respect to the lowest loss value are shown in Eq.~\ref{sr_results}.a-c with one, two and all four of the variables $\cos\theta_{1-4}$, respectively.
\begin{subequations}
\label{sr_results}
\begin{align}
    f_{q,3}(\theta_4)               
    &= a\cdot \cos \theta_4 \\
    f_{q,5}(\theta_1, \theta_2)     
    &= b\cdot \left(\cos \theta_1  + \cos\theta_2 \right)\\
    f_{q,11}(\theta_1, \theta_2, \theta_3, \theta_4)
    &=  \left( c\cdot \cos\theta_1 + \cos \theta_2 - \cos\theta_3\right) \cdot \left(\cos\theta_4 + d \right) 
\end{align}
\end{subequations}
The parameters are $a=6.22$, $b=-5.76$, $c=-1.51$ and $d=4.63$. 
Fig.~\ref{fig:NN_representation}.d tests how well these analytical models of the logit committor reproduce the sampled logit committor.

The model with the lowest complexity that is still valid, Eq.~\ref{sr_results}.a, corresponds to a scaled single input variable $\cos\theta_4$. Input variable $\cos\theta_3$ could be also used for a valid logit committor function but with worse accuracy according to the loss function.
Eq.~\ref{sr_results}.b is the first model with two input variables and corresponds to a scaled sum of $\cos\theta_1$ and $\cos\theta_2$. Some SR runs also suggest a scaled sum of $\cos\theta_3$ and $\cos\theta_4$ but with a $0.1\%$ higher loss value. 
The pairing of the dihedral angles is noteworthy:
$\theta_1$ and $\theta_3$ consist of heavy atoms and contain C$_{12}$.
Since $\theta_1$ and, to a lesser degree, $\theta_3$ represent the main chain movement, it is understandable that the model assigns the highest significance to these to dihedrals. 
$\theta_2$ and $\theta_4$ both contain the hydrogen substituent at C$_{14}$ and have a lower significance.
In the analytical models with two input variables, a high-significance dihedral is paired with a low-significance dihedral, such that all four substituents to the C$_{13}$=C$_{14}$ double bond are represented.

Eq.~\ref{sr_results}.c is a model which combines all four input variables. 
In contrast to Eq.~\ref{sr_results}.a and \ref{sr_results}.b, it is a non-linear function of the input variables.
Its loss value is about $8\%$ lower than Eq.~\ref{sr_results}.b.
Even though this decrease in loss seems modest in Fig.~\ref{fig:NN_representation}.c, in the visual comparison between predicted and sampled logit committor, Eq.~\ref{sr_results}.c is the only analytical model that reaches comparable accuracy to the reference NN model. SR shows that, even though a 2-dimensional model of the committor is qualitatively correct, a 4-dimensional non-linear model is required to be quantitatively accurate. 
In total, we have generated two NN models and three analytical models of the logit committor. 
Pairwise comparison of these models and the comparison to the sampled logit committor are presented in Fig.~\ref{fig:logit_committor_comparison}.
        
\subsection{Reaction mechanism}
    \begin{figure}[htb]
        \centering
        \includegraphics[width=1.0\linewidth]{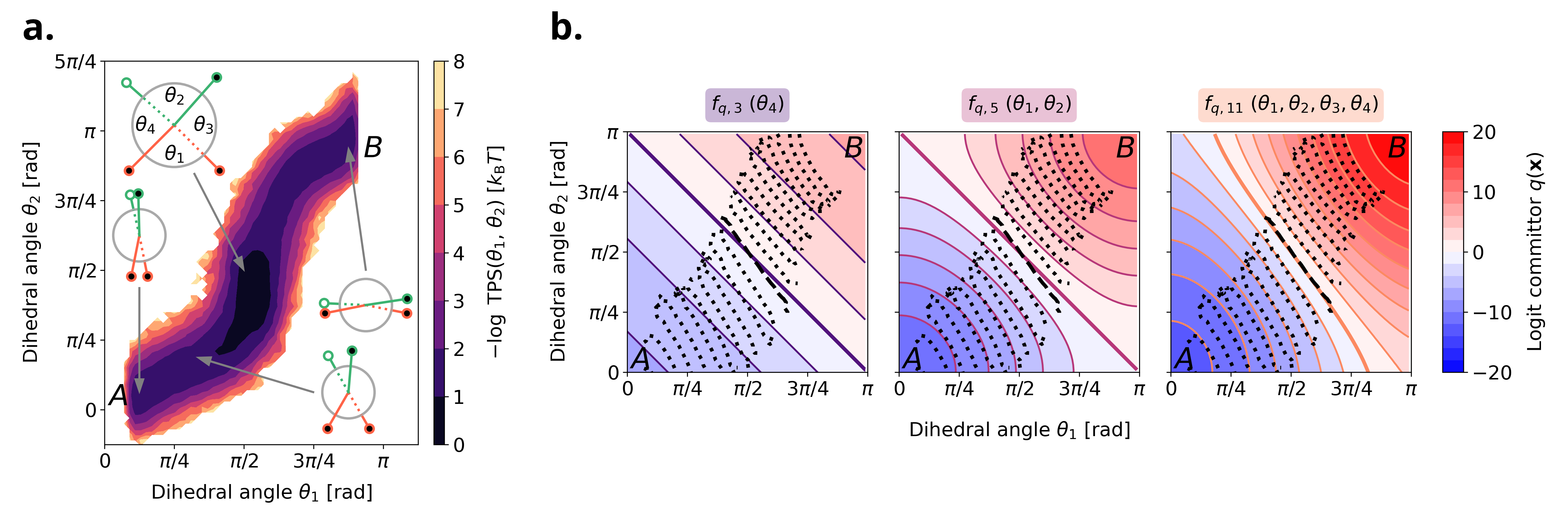}
        \caption{
        {\bf (a.)} TPE density profile in the $\theta_1, \theta_2$ dihedral angle space
        with sketches of Newman projections of all 4 $\theta_i$ angles labeled in the projection of the TS (top left).
        {\bf (b.)} Isocontour lines of the logit committor $q(\mathbf{x})$ projected in $\theta_1, \theta_2$ 
        dihedral angle space with $\theta_3 = \theta_4 = \pi - (\theta_1 + \theta_2)/2$
        as $\sum_{i=1}^4 \theta_i = 2\pi$.
        The colormap and colored solid lines depict the isocontour lines of the logit committor
        according to the SR results with complexity 3, 5 and 11, respectively.
        The black dashed and dotted lines are the isocontour lines of the average logit committor values
        predicted by the NN logit committor model based on internal coordinates 
        from conformations within the respective ($\theta_1, \theta_2$) bins.
        All lines depict a logit committor difference of $\Delta q_B = \pm2$ 
        and the bold colored lines and black dashed lines indicate the logit committor of $q_B = 0$.
        }
        \label{fig:srres}
    \end{figure}

Fig.~\ref{fig:srres}.a shows the same TPE as in Fig.~\ref{fig:problemStatement}.c but now projected into the space of the dihedrals $\theta_1$ and $\theta_2$.
The conformations of the two dihedrals are shown as Newman projections ($\theta_1$=orange, $\theta_2$=green) for selected points along the reaction. 
Carbon atoms are marked with black filled circles.
In this projection, the zig-zag motion in Fig.~\ref{fig:problemStatement}.c is transformed into an S-curved path which predominantly  proceeds via $\theta_1$ in regions close to the reactant and product states but via $\theta_2$ around the TS.
This step-wise mechanism generates the out-of-plane bending at the carbon atoms, since $\theta_1$ moves from 0 to almost $\pi/2$ while $\theta_2$ largely stays fixed. 
At the TS both dihedrals assume an angle of $\pi/2$.
The NN model based on internal coordinates (Fig.~\ref{fig:AIMMD}.b) captures this S-curved reaction path, 
as demonstrated by the isocontours of the (average) logit committor in Fig.~\ref{fig:srres}.b (black dots).
Thus, despite the high reaction barrier, AIMMD successfully models details of the reaction mechanism beyond the TS ensemble at $p_B(\mathbf{x})=0.5$.
Fig.~\ref{fig:srres}.b also shows how accurately the analytical models (Eq.~\ref{sr_results}.a-c) reproduce the underlying (average) NN model predictions.
While $f_{q,5}(\theta_1, \theta_2)$ (Eq.~\ref{sr_results}.b) is already a function of the $(\theta_1, \theta_2)$ space, the other two models need to be slightly transformed to be represented in this space. 
For that, we use that the four dihedral angles sum up to a full circle ($\sum_{i=1}^4 \theta_i = 2\pi$). 
This is evident from the Newman projection of the TS in Fig.~\ref{fig:srres}.a.
From the other Newman projections in Fig.~\ref{fig:srres}.a, it is also evident that the opposing dihedrals approximately have the same value, i.e. $\theta_3 \approx \theta_4$. 
Assuming that they have \emph{exactly} the same value, we can assign their value in
$f_{q,11}(\theta_1, \theta_2, \theta_3, \theta_4)$ as
\begin{equation}
    \label{eq:theta_relation}
    \theta_3 = \theta_4 = \pi - (\theta_1 + \theta_2)/2 ~~~.
\end{equation}
The same relation allows us to convert any point in the $(\theta_1, \theta_2)$-space into a value of $\theta_4$ and insert it into $f_{q,3}(\theta_4)$ (Eq.~\ref{sr_results}.a). 
The isocontour lines of the resulting functions are shown in Fig.~\ref{fig:srres}.b.
In $f_{q,3}(\theta_4)$ and $f_{q,5}(\theta_1, \theta_2)$, the reaction mechanism, defined by the gradient vectors orthogonal to the isocontour lines, follows a straight line connecting the \textsl{cis}-state at $(0,0)$ and the \textsl{trans}-state at $(\pi,\pi)$.
In doing so, these two analytical models fail to capture the S-shaped reaction pathway of the NN model.
Even though the TS ensemble at $p_B(\mathbf{x})=0.5$ (i.e.~$f_{q,3}(\theta_4)= f_{q,5}(\theta_1, \theta_2)=0$) is well-aligned with the NN model predictions, the analytical approximations $f_{q,3}(\theta_4)$ and $f_{q,5}(\theta_1, \theta_2)$ smooth out the crucial feature that the reaction proceeds via a step-wise mechanism.
$f_{q,5}(\theta_1, \theta_2)$ improves on $f_{q,3}(\theta_4)$ by capturing that the state boundaries of reactant and product states are approximately circular.
Only the model that uses all four dihedral angles identified by the HIPR analysis, $f_{q,11}(\theta_1, \theta_2, \theta_3, \theta_4)$, accurately captures the S-curvature of the reaction pathways. 
In this analytical form, the cosine terms are coupled multiplicatively (Eq.~\ref{sr_results}.c).
This nonlinear coupling induces isocontour lines with an asymmetric curvature relative to the diagonal of the $(\theta_1, \theta_2)$-space.
These isocontour lines perfectly align with the NN model, showing that the NN model can in fact be represented by a low-dimensional analytical expression.

\section{Origin of the reaction mechanism}

\begin{figure}[htb]
\centering
\includegraphics[width=1\linewidth]{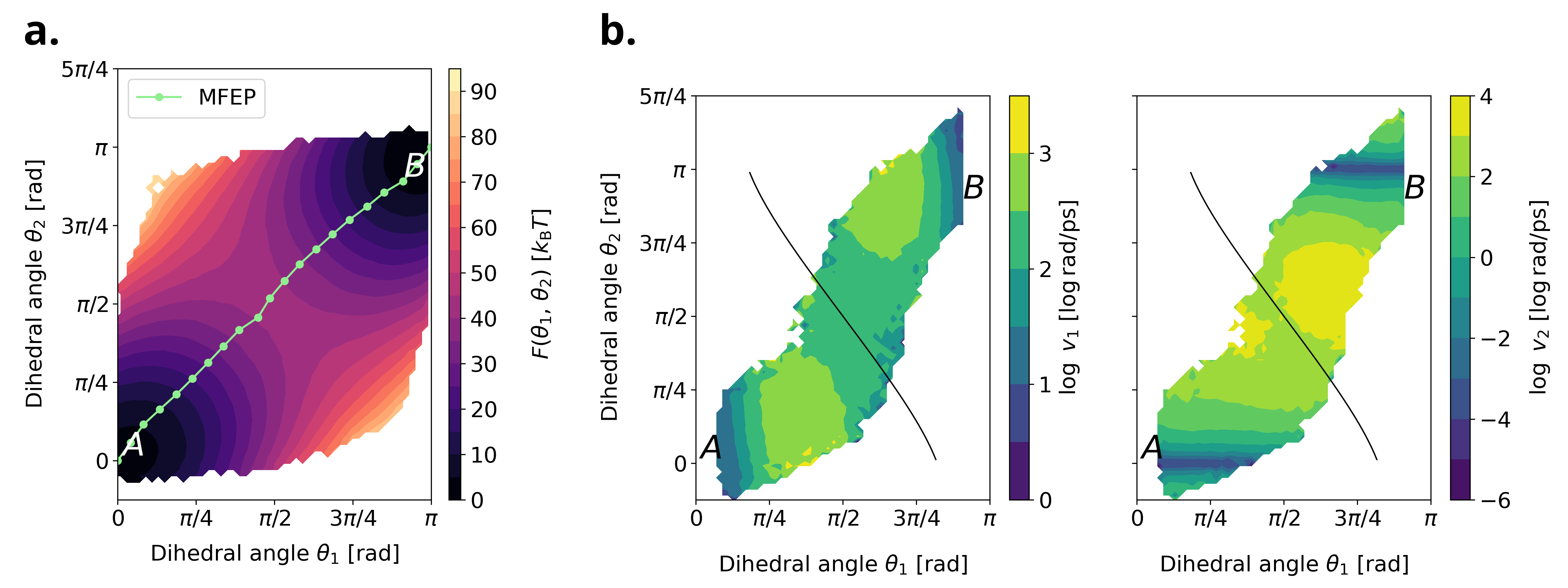}
\caption{
    {\bf (a.)} FES from a series of umbrella simulation and the minimum free energy path (MFEP, dotted light green line) from NEB on the FES between ($A$) \textsl{cis} and ($B$) \textsl{trans} configuration.
    {\bf (b.)} TPE average dihedral angle velocities (left) $v_1 = \langle\dot\theta_1\rangle_\mathrm{TPE}$ and (right) $v_2 = \langle\dot\theta_2\rangle_\mathrm{TPE}$ in frames with the respective ($\theta_1, \theta_2$) configurations. The solid black line marks the zero logit committor estimated by Eq.~\ref{sr_results}.c.
    Note the different color scale boundaries.
}
\label{fig:TPE}
\end{figure}

We have established that the reaction follows an S-shaped pathway with a stepwise mechanism, associated with out-of-plane bending of the carbon atoms.
The origin of this mechanism, however, remains unclear. 
This is particularly puzzling because the free-energy surface in the $(\theta_1, \theta_2)$-space space does not exhibit an S-shaped profile. 
Instead, the minimum free-energy path follows a straight line between reactant and product states.
This contrasts with the TPE and suggests that the S-shaped reaction pathway does not arise from the underlying potential energy surface, but emerges from the dynamics.
Fig.~\ref{fig:TPE}.b shows the average dihedral angle velocities 
$v_1 = \langle\dot{\theta}_1\rangle_{\mathrm{TPE}}$
and $v_2 = \langle\dot{\theta}_2\rangle_{\mathrm{TPE}}$ averaged over the TPE.
The graphs highlight that, although both angle geometrically describe the same torsion around the C$_{13}$=C$_{14}$ double bond, they are not equivalent from the viewpoint of the reaction mechanism.
First, the reaction mechanism is asymmetric at the configurational level: the system evolves predominantly along $\theta_1 \rightarrow \theta_2 \rightarrow \theta_1$, whereas the alternative sequence $\theta_2 \rightarrow \theta_1 \rightarrow \theta_2$ would correspond to a different reaction mechanism. 
Second, the velocity profiles of the two dihedral angles differ.
$\langle\dot{\theta}_1\rangle_{\mathrm{TPE}}$ reaches a maximum on either side of the TS, but $\theta_1$ is relatively slow at the TS itself. 
By contrast, $\theta_2$ has overall higher velocities and reaches its maximum velocity in the immediate vicinity of the TS.

We interpret these results as follows. The free-energy surface represents a fully thermalized canonical ensemble at each grid point. 
Likewise, the two-way shooting simulations sample the canonical ensemble, as initial velocities are drawn from the Maxwell–Boltzmann distribution and a thermostat is applied throughout. 
The ergodicity theorem relates the canonical stationary distribution to the underlying path ensemble, comprising both rare-event excursion and reactive paths.
Partitioning the path ensemble into rare-event excursions and reactive trajectories splits the velocity distribution unevenly, giving rise to the position-dependent velocity profiles in Fig.~\ref{fig:TPE}.b.
Neither of the two subensembles is fully thermalized. 
Since the TPE is much smaller than the ensemble of non-reactive trajectories, deviations from the stationary distribution are more pronounced in the TPE. 
In this conditional ensemble, configurations near the transition state are overrepresented compared to equilibrium.
The strong forces acting on the atoms near the reaction center rapidly convert kinetic into potential energy (and vice versa) within the reactive degrees of freedom.
Because the transition paths are very short ($0.128\pm0.054$\,ps on average), this energy redistribution cannot equilibrate with the surrounding degrees of freedom or the thermostat. 
This inertial barrier-crossing dynamics generate the position-dependent velocity bias in the TPE.
Finally, the asymmetry between $\theta_1$ and $\theta_2$ arises from different effective masses associated to these degrees of freedom. 
While $\theta_1$ is defined in terms of four heavy atoms, $\theta_2$ contains the hydrogen substituent at C$_{14}$. 
$\theta_1$ therefore has a higher effective mass than $\theta_2$.
Visual inspection of individual transition paths reveals pronounced oscillations around the minimum-energy path, indicating that the TPE should be interpreted in an average sense. 
With this caveat in mind, the velocity profiles suggest the following dynamical picture: the reaction is initiated by an excursion from the \textsl{cis}-state primarily along the heavier degree of freedom $\theta_1$, during which kinetic energy is converted into potential energy as $\theta_1$ approaches the transition region. 
The resulting strained, out-of-plane conformation is then relieved by a rapid transition of the hydrogen at C$_{14}$, which generates the zig-zag pattern in Fig.~\ref{fig:problemStatement}.c. 
This release of potential energy likely transfers kinetic energy back to $\theta_1$, facilitating barrier crossing and thereby completing the transition to the \textsl{trans} state.

\section{Conclusions}

Using the \textsl{cis}--\textsl{trans} isomerization of retinal as a test case, 
we have demonstrated that AIMMD can efficiently sample a high-barrier chemical 
reaction and learn its committor accurately --- not only at the transition-state 
ensemble $p_B(\mathbf{x})=0.5$, but well into the reactive region on either side. 
This is a non-trivial achievement: for barriers as high as that of retinal 
isomerization (${\sim}100$\,kJ/mol in our force field), the committor is 
effectively a step function, and a model that merely identifies the 
isocommittor surface at $0.5$ would appear accurate on most test points 
while missing the mechanistic information that lives in the narrow interval 
around it. AIMMD avoids this failure mode by actively sampling configurations 
across the full logit-committor range and by learning $q_B(\mathbf{x})$ rather 
than $p_B(\mathbf{x})$ directly, since the logit varies smoothly where the 
committor saturates; the committor itself is then obtained straightforwardly via the inverse logit. Thorough path-ensemble sampling and a smooth learning 
target together enable the model to resolve even the small-amplitude motions 
that shape the pathway.
The choice of coordinates through which the committor is expressed proved 
decisive. When viewed through the improper dihedrals at C$_{13}$ and C$_{14}$, 
the small-amplitude motions appear as a flip--flop: reactant, transition, and 
product states all share $\chi_1 = \chi_2 = 0$, so these coordinates carry 
little information about reaction progress and are unsuitable for 
parametrizing the logit committor.
HIPR analysis revealed that the neural network instead relies on the four 
proper dihedrals around the C$_{13}$=C$_{14}$ bond. In this representation, 
the pathway becomes monotonic from \textsl{cis} to \textsl{trans}, and what 
looks like a flip--flop in $(\chi_1,\chi_2)$ is recognized as a fundamentally 
stepwise mechanism in $(\theta_1,\theta_2)$.
Symbolic regression then distilled the neural-network representation into 
analytical expressions of the logit committor. Compared to the NN, these 
expressions are human-interpretable and computationally cheap, making them 
suitable as collective variables in biased simulations \cite{france2024data} 
or for analytical rate estimates \cite{bolhuis2002transition,vanden2010transition}. 
They also extrapolate smoothly beyond the sampled region, where NN predictions 
become noisy --- though the accuracy of such extrapolations, in the absence 
of sampling to check them against, remains an open question.
A cautionary finding emerged from this procedure: the loss is not a reliable 
indicator of whether an analytical model captures the mechanism. Moving from 
two to four input coordinates reduces the loss by only 8\%, yet only the 
four-coordinate model reproduces the S-shaped reaction path; the 
two-coordinate model smooths it into a straight line connecting reactant 
and product. Chemically meaningful improvements are concentrated in the tail 
of the learning curve, which complicates overfitting control and model 
selection in symbolic regression for reaction-coordinate discovery.
Perhaps the most striking observation of this work is that the free-energy 
landscape is, on its own, a misleading guide to the reaction mechanism. 
The minimum-free-energy path in $(\theta_1,\theta_2)$ is a straight diagonal 
connecting \textsl{cis} and \textsl{trans}, suggesting a concerted torsion 
around the C$_{13}$=C$_{14}$ bond. The transition-path ensemble tells a 
different story: reactive trajectories follow an S-shaped, stepwise path 
dominated by $\theta_1$ near the basins and by $\theta_2$ near the transition 
state. The discrepancy is not a failure of the free-energy calculation but a 
consequence of what the two objects represent. The FES integrates over all 
configurations at a given $(\theta_1,\theta_2)$, including the overwhelming 
majority that belong to non-reactive fluctuations; the reactive density is 
masked by this non-reactive background. The TPE, by contrast, is conditioned 
on reactivity and therefore isolates the flow that actually produces the 
transition.
A second reason for the discrepancy is dynamical. The TPE is not a canonical 
ensemble: transition paths are too short ($0.128 \pm 0.054$\,ps on average) 
for the strong forces near the reaction center to thermalize with the 
surrounding degrees of freedom. Combined with the mass asymmetry between 
the heavy-atom dihedral $\theta_1$ and the hydrogen-bearing $\theta_2$, this 
non-equilibrium regime produces the asymmetric 
$\theta_1 \rightarrow \theta_2 \rightarrow \theta_1$ sequence that underlies 
the mechanism. The reaction mechanism is, in this sense, a dynamical 
phenomenon that neither the MEP nor the FES can reveal on its own.
This echoes an insight from gas-phase reaction dynamics: even when the potential energy surface is well characterized, it does not uniquely determine the reaction mechanism, which is governed by the underlying dynamics \cite{polanyi1987some, miller1993beyond, herschbach1987molecular}.

This observation also exposes a limitation of the current AIMMD protocol, 
which incorporates equilibrium assumptions at several stages: shooting points are selected using the overdamped formulation of $P(\mathrm{TS} | \mathbf{x})$ and, more 
importantly, the logit committor is formulated solely in terms of positions, 
with velocities drawn from the Maxwell-Boltzmann distribution at shooting. 
Underdamped dynamics increases the region in which $P(\mathrm{TS} | \mathbf{x})$ is high\cite{brunig2025nonmarkovianity}. But since this probability is employed here solely as a sampling heuristic, it does not contribute to the loss function and therefore has no influence on the learned neural network representation of the logit committor.
By constrast, extending the framework to the phase-space committor 
\cite{bolhuis2002transition,vanden2010transition,roux2021string} is 
a natural next step, and one likely to matter more for high-barrier reactions 
than for the diffusive processes where AIMMD has previously been applied. 
A second extension follows from AIMMD's reliance on transition path sampling, 
which naturally assumes a two-state description: multistate systems can be 
addressed by sampling transitions between multiple pairs of metastable 
states, for example within a core-set Markov model approach 
\cite{buchete2008coarse,schutte2011markov,lemke2016density} or through 
multiple-state TPS \cite{rogal2008multiple}.
Importantly, AIMMD can readily be applied to larger and more flexible molecules than retinal and to reactions which require an atomistically modelled environment.

Taken together, our results show that learning the logit committor with 
neural networks, interpreting it with HIPR, and distilling it with symbolic 
regression offers a coherent route from unbiased molecular dynamics to a 
human-readable reaction coordinate. For retinal isomerization, this route 
recovers the torsional mechanism and reveals its origin: the mechanism is 
shaped not by the energy landscape alone, but by the non-equilibrium, 
reactive dynamics that unfold on it --- dynamics to which the free-energy 
surface is blind.

%
%
\newpage 
\section{Methods}

\subsection{Molecule and force field}
Object of this investigation is the positively charged retinal-lysine Schiff base with the lysine saturated by an amide bond with acetic acid (N-terminus) and methylamine (C-terminus) shown in Fig.~\ref{fig:problemStatement}b.
The \textsl{cis}-\textsl{trans} isomerization reaction is defined by the change of dihedral angles $\theta_i$ around the C13-C14 bond.
The potential energy surface of the molecule is provided by a modified AMBER99SB*-ILDN force field\cite{ll2010ff99sb} where the parameters of the retinal are adopted to GROMACS format\cite{malmerberg2011ambermod,hub2011ambermod} from the work of Hayashi and coworker.\cite{hayashi2002ambermod}
The initial molecular structure for retinal-lysine were obtained from experimentally measured crystal structure of channelrhodopsin 2.\cite{volkov2017xray}

All MD simulations in AIMMD are carried out at 300\,K in vacuum with GROMACS version 2018.8 (and 2019.6 for umbrella simulations)\cite{spoel2005gromacs, abraham2015gromacs} using the same settings as in Ref.~\citenum{ghysbrecht2025accuracy}.
We simulated the system using the leap-frog stochastic dynamics integrator with a 2\,fs timestep and an inverse friction coefficient of 2\,ps.
The LINCS constraint algorithm was applied to constrain all hydrogen bonds.\cite{hess1997lincs}

\subsection{Minimum energy path}

The minimum energy path (MEP) between the equilibrated \textsl{cis} and \textsl{trans} configuration of the retinal-lysine molecule is predicted by the climbing image nudged elastic band (ci-NEB) method implemented in the atomic simulation environment (ASE) version 3.25.0.\cite{larsen2017ase,henkelman2000cineb}
The same set of AMBER force field parameters are used for the energy and force evaluation computed by its implementation in OpenMM version 8.2 and interfaced to ASE using the Narupa Python package.\cite{eastman2024openmm,oconnor2019narupa}
    
\subsection{AIMMD}

We used AIMMD for generating trajectory data while iteratively learning an initial NN representation of the logit committor. Then, we trained a final NN representation on the full training set. Finally, we used HIPR and symbolic regression to derive  analytical formulas that serve as an approximation of the logit committor.

\paragraph{Data production.}
    The simulations focus on the \textsl{cis}-\textsl{trans} isomerization around the $\mathrm{C}_{13}=\mathrm{C}_{14}$ double bond of retinal, covalently coupled to a lysine residue. 
    To characterize the system, we selected the dihedral angle $\theta_1 \in (-\pi, \pi]$ defined by sp\textsuperscript{2}--carbons 12, 13, 14, and 15 (see Fig.~\ref{fig:problemStatement}). 
    $\theta_1$ describes the isomerization process as expressed in the state definition function 
    \begin{equation}
    \text{SDF}(\theta_1) =
    \begin{cases}
        \text{A}, & \text{if } \left| \theta_1 \right| < \frac{\pi}{9} \quad \quad \quad\text{(cis)} \\[8pt]
        \text{B}, & \text{if } \left| \theta_1 \right| > \pi - \frac{\pi}{9} \quad \text{(trans)} \\[8pt]
        \text{R}, & \text{otherwise} \quad \quad \quad \text{(reactive)}
    \end{cases}
    \label{eq:state_definition_function}
    \end{equation}
    
    AIMMD sampled the path ensemble that represents the retinal isomerization event from $A$ via the TS at $\theta_1 \sim \pi/2$ to $B$. (The opposite rotation was not considered.) 
    The algorithm requires an initial transition path (TP) to initialize the early simulations; this path does not need to represent a physically accurate trajectory. 
    We generated the initial path from two-way shooting from a high energy configuration with $\theta_1 \sim \pi/2$ and assumed to be at the TS.
    AIMMD generates paths by performing two-way shooting from previous trajectories. Simultaneously, it updates a NN as an intermediate step (Fig.~\ref{fig:AIMMD}.a), which in turn refines the shooting point selection based on the most recent NN model.

    During the data production process, the NN model is a sub-optimal approximation of reaction coordinate model $q_B({\bf x})$.
    The model for the generation process (Fig.~\ref{fig:AIMMD}.a) employed a feedforward neural network (FFNN) with 3,160 input features corresponding to pairwise atom-atom distances. The model architecture is composed of three layers with PReLU activation functions. We employ ADAM\cite{ba2017adam} for optimization with an initial learning rate equal to $l_r  = 10^{-5}$.
    At regular AIMMD steps, following each completed two-way shooting simulation, we retrain the networks from scratch using a loss function derived from the simulation outcomes (Eq.~\ref{eq:loss_function}). The training set matches the number of simulated excursions and transitions, containing $({\bf x}_{i}, r_{\text{A},i}, r_{\text{B},i})$ triplets. Here, ${\bf x}_i$ denotes the shooting points atom positions of the path, while $r_{\text{A},i}$ and $r_{\text{B},i}$ represent the counts of boundary crossings for states A and B, respectively. Subsequently, we minimize the loss function in Eq.~\ref{eq:loss_function}
    over 500 training epochs, resampling the training set without replacement in batches of size 4096.
    
    We perform path sampling excluding equilibrium simulations. The sampling strategy and reweighting procedure ensure a uniform distribution in the $p_B({\bf x}) \approx \sigma(q_B({\bf x}))$ space, where $p_B$ is an optimal reaction coordinate that defines the transitions between two states, between the manually defined interfaces $q_{B,\mathrm{min}}=-20$ and $q_{B,\mathrm{max}}=+20$.
    These interface values place the system sufficiently close to the states while avoiding direct proximity to the boundaries.
    These settings enable the system to sample the shooting point from a broader range of paths, rather than relying solely on the most recent path, which eventually ensures more extensive exploration of the reaction space and prevents stagnation near the states. 
    This procedure may restrict exploration in directions orthogonal to the committor. Our simulations show that within the range of positive $\theta_1$ values, only one pathway exists, indicating rapid equilibration in that direction.
    Lazzeri et al. provide additional settings and detailed guidance on the methodology employed.\cite{lazzeri2023aimmd}


    The final dataset consists of a total of 32,727 atomistic simulations representing shooting paths, with 9.1\% classified as transitions and the remaining categorized as excursions around the state basin.

    To project the transition path ensemble density profile, we reweight the paths to ensure they populate the target ensemble in equilibrium proportions. The selection of shooting points follows the Boltzmann distribution along the isocommittor surfaces, as guaranteed by the theoretical framework underlying sampling in AIMMD \cite{lazzeri2025optimal}.
    We assign a reweighting factor to the paths, proportional to the inverse of the number of crossings at the shooting interface. Here, we use an approach consistent with established methodologies \cite{falkner2023conditioning,best2005reaction}, where the reweighting ensures proper Boltzmann-weighted sampling.
    After reweighting, AIMMD sampling becomes equivalent to transition interface sampling, with the interfaces aligned to the locations of the shooting points. This step precedes the path ensemble reweighting procedure described in detail by Lazzeri \cite{lazzeri2023aimmd}.


\paragraph{NN embedding of the committor}
    In general, it is always possible to improve the model after the sampling campaign has ended \cite{jackel2025free}. 
    We retrained a custom FFNN implemented in PyTorch,\cite{paszke2019pytorch} comprising three fully connected layers with a total of 797,185 parameters. 
    The network consists of an input layer, two hidden layers, each containing 512 nodes, and a one-node output layer. A dropout rate of 0.5 is applied after the input layer and the first hidden layer. After each layer, the network uses a PReLU activation function. We employed the ADAM optimizer with an initial learning rate of $10^{-5}$, training the model for 500 epochs with a batch size of 32.
    The input features were derived from the internal coordinate representation of the shooting points, including 80 bond distances, 144 angles, and 157 dihedral angles projected as sine and cosine components, resulting in a total of 528 features.
    The shooting results $\mathbf{r}_i=(r_{i,A}, r_{i,B})$ were defined from the outcomes of the two-way shooting initiated at $\mathbf{x}_i$, capturing the frequency with which each path visits states A and B. We adopted the loss function used within the AIMMD framework and the same rewriting strategy.
    We split the dataset based on the AIMMD production runs, with a ratio of 77:23 for training and testing, respectively. This division ensured that the correlation between the test and train dataset was minimal.
    
    We validated and assessed the accuracy of the committor learned by the NN using a dedicated validation set consisting of 482 shooting points excluded from prior steps. For each data point in the validation set, we performed a two-way shooting procedure, yielding 400 shooting outcomes, where $N_A$ is the number of shooting outcomes that reach $A$ before $B$ and $N_B$ is the number of outcomes that reach $B$ before $A$. 
    The committor estimate $N_B/(N_A + N_B)$ is compared to the neural network prediction in Fig.~\ref{fig:NN_representation}.a / Fig.~\ref{fig:logit_committor_comparison} that shows a strongly aligned estimation with the model prediction. 

\paragraph{Symbolic regression}
    
    We then performed symbolic regression using the PySR framework \cite{cranmerInterpretableMachineLearning2023}.
    We injected the four most significant features identified through HIPR analysis as inputs for the symbolic regression, and we used the two-way shooting results as labels. To maintain methodological consistency, we applied the same AIMMD loss function to this optimization problem and preserved the same dataset split into training and testing subsets.
    The symbolic regression framework permitted algebraic sum, multiplication, division, logarithm, and exponential functions, constraining the maximum complexity to 20 and maximum depth to 7. We performed 1000 iteration for each cycle. All other parameters remained at their default values.
    
    To mitigate stochastic effects arising from system initialization, we executed three independent symbolic regression runs using identical parameters. We assessed the trade-off between model complexity and performance through validation set evaluations, ensuring a robust analysis of the derived analytical expressions.

\subsection{Free-energy surface}

The free energy surfaces in this work are computed via umbrella sampling simulations at different bias windows and the weighted histogram analysis method (WHAM) \cite{kumar1992wham,souaille2001wham}.
The umbrella sampling simulations are performed using the same MD simulation setup with GROMACS version 2019.6 together with the PLUMED plugin \cite{tribello2014plumed}.
The simulations also applies the leap-frog stochastic dynamics integrator with an inverse friction coefficient of 2\,ps but a 1\,fs timestep.
The CV is chosen according to the results of the symbolic regression to obtain an approximate expression of the committor function.
For each umbrella simulation a harmonic bias with a force constant of 300\,kJ/mol/$\Delta q^2$ is applied, where $\Delta q$ is the spatial deviation from the respective point along the CV.

The MFEP is computed by the NEB method with 21 beats and spring force of\\2\,kJ/mol/\AA~between the \textsl{cis} state $A$ ($\theta_1=0$, $\theta_2=0$) and the the \textsl{trans} state $B$ ($\theta_1=\pi$, $\theta_2=\pi$) using the 2-dimensional FES.

\subsection{TPE velocity}

The dihedral angle velocities in the transition path ensemble are computed from the atom velocities in Cartesian coordinates ${\partial {\bf x}} / {\partial t}$ stored in the AIMMD-generated trr trajectory files. The partial derivative ${\partial \theta} / {\partial {\bf x}}$ was determined to convert from Cartesian atom velocities to angle velocity by:
\begin{equation}
    \dfrac{\partial \theta}{\partial t} = \dfrac{\partial \theta}{\partial {\bf x}} \dfrac{\partial {\bf x}}{\partial t}
\end{equation}
The projection of the dihedral angle velocities of the TPE along certain angles show the average of the absolute dihedral angle velocities of all frames with retinal-lysine configuration of the respective projected angles bin.

%
%
%

\begin{acknowledgement}
This research has been funded by the Volkswagen Foundation through a Momentum grant. 
This work was supported by Goethe University Frankfurt (G.L., V.O., and R.C.); the Frankfurt Institute of Advanced Studies (G.L., V.O., and R.C.); the 
Quandt Foundation (G.L., V.O., and R.C.); the German Research Foundation, CRC 1507: Membrane-Associated Protein Assemblies, Machineries, and Supercomplexes (P09) (project id: 450648163) (G.L., V.O., and R.C.);  the Center for Scientific Computing of the Goethe University (G.L., V.O., and R.C.); the Juelich
Supercomputing centre (G.L., V.O., and R.C.).
\end{acknowledgement}

\newpage
\begin{suppinfo}

\subsection{Log-likelihood of a Bernoulli-trial}
For $k$ independent Bernoulli trials, where each trial has its own probability $p_B(\mathbf{x}_i | \mathbf{w})$, the negative log likelihood is
\begin{align}
    -\ln P(S|\mathbf{w}) 
    &= -\ln \prod_{i=1}^k p_B(\mathbf{x}_i|\mathbf{w})^{\delta_i} \cdot (1-p_B(\mathbf{x}_i|\mathbf{w}))^{1-\delta_i} \cr
    &= \sum_{i=1}^k \delta_i \left(-\ln  p_B(\mathbf{x}_i|\mathbf{w})\right) + (1-\delta_i) \left( -\ln(1-p_B(\mathbf{x}_i|\mathbf{w})) \right)\cr
    &= \sum_{i=1}^k \delta_i \ln (1+e^{-q(\mathbf{x}_i|\mathbf{w})}) + (1-\delta_i) \ln (1+e^{+q(\mathbf{x}_i|\mathbf{w})})\cr
    &= \sum_{i=1}^k \delta_i \ln (1+e^{s_iq(\mathbf{x}_i|\mathbf{w})})+ (1-\delta_i) \ln (1+e^{s_iq(\mathbf{x}_i|\mathbf{w})})\cr
    &= \sum_{i=1}^k \ln(1+e^{s_iq(\mathbf{x}_i|\mathbf{w})})
\label{eq:logLikelihood}    
\end{align}
where $\delta_i$ is a binary outcome variable with $\delta_i=1$ if the trajectory reaches $B$ before $A$, and $\delta_i=0$ otherwise. 
The signed outcome variable is $s_i = -1$ if trajectory $i$ reaches $B$ before $A$, and $s_i = +1$ if it reaches $A$ before $B$.
The two outcome variables are related by $s_i = 1 - 2\delta_i$.
In Eq.~\ref{eq:logLikelihood} we used the identities in Eq.~\ref{eq:logit_committor}
\begin{align}
    -\ln p_B    &= - \ln \frac{1}{1+e^{-q}} = \ln (1+e^{-q}) \cr
    -\ln(1-p_B) &= - \ln \left( 1-  \frac{1}{1+e^{-q}} \right) 
                 = - \ln \left( \frac{e^{-q}}{1+e^{-q}} \right)
                 = \ln \left( \frac{1+e^{-q}}{e^{-q}} \right)
                 = \ln (1+e^{+q}) \, .
\end{align}
Here, we abbreviated $p_B(\mathbf{x}) = p_B$ and $q_B(\mathbf{x}) = q$.

\subsection{Figures}
\begin{figure}
    \centering
    \includegraphics[width=1\linewidth]{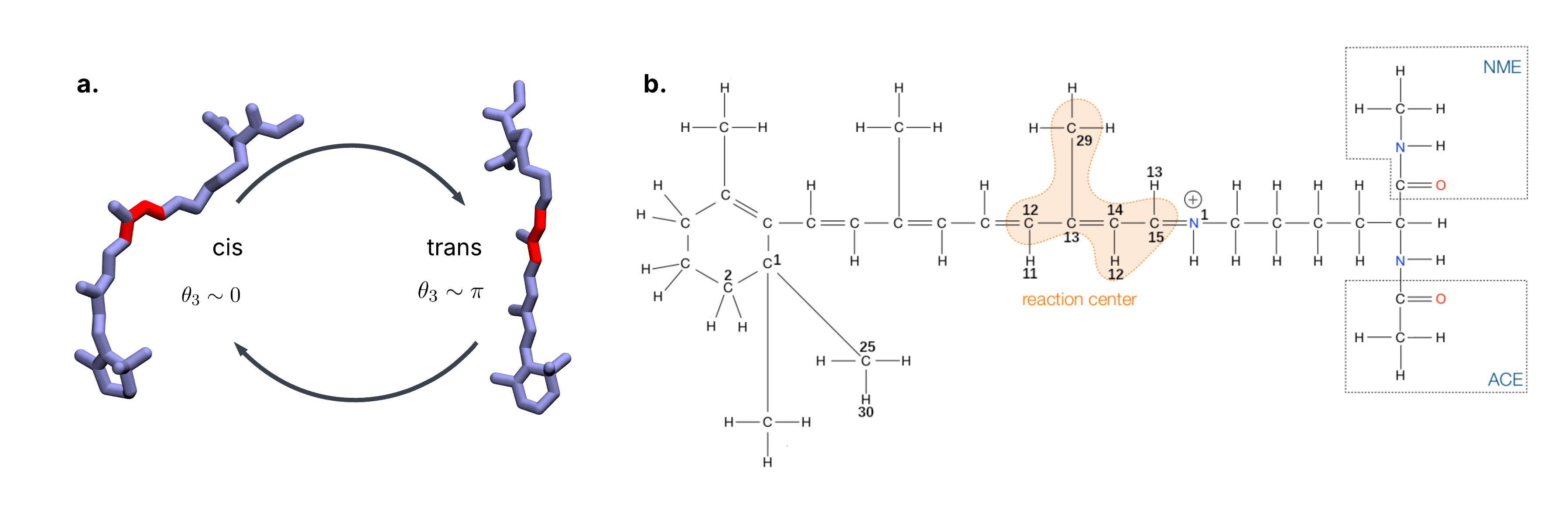}
    \caption{{\bf (a.)} \textsl{cis}-\textsl{trans} isomerization of N-retinylidene-lysine
    {\bf (b.)} Topology of the molecule with reaction center highlighted.}
\label{fig:reaction_center}
\end{figure}
    
\begin{figure}
    \centering
    \includegraphics[width=0.75\linewidth]{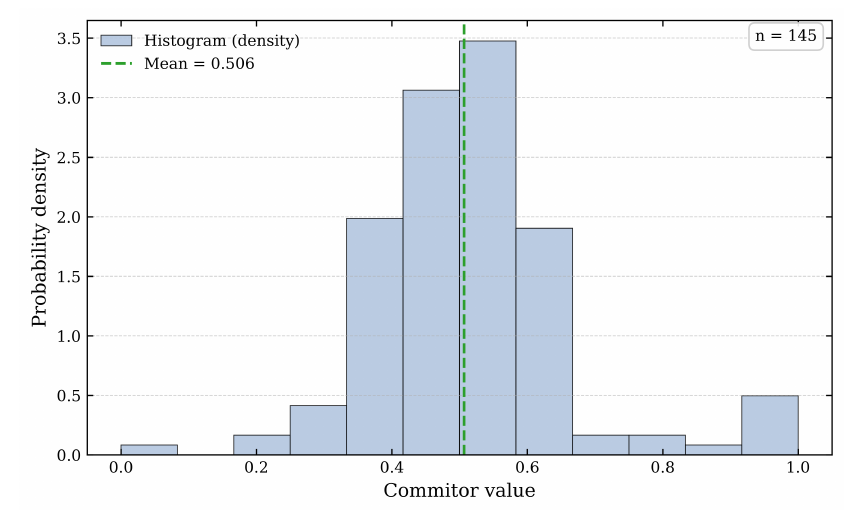}
    \caption{ 
    A total of 145 retinal conformations with predicted committor values near $0.5$ were sampled. Their corresponding estimated committor values were obtained from two-way shooting simulations. The resulting distribution closely follows a normal distribution, with a mean committor value of $0.506$.}
    \label{fig:halfcmt}
\end{figure}
    
\begin{figure}
    \centering
    \includegraphics[width=1.0\linewidth]{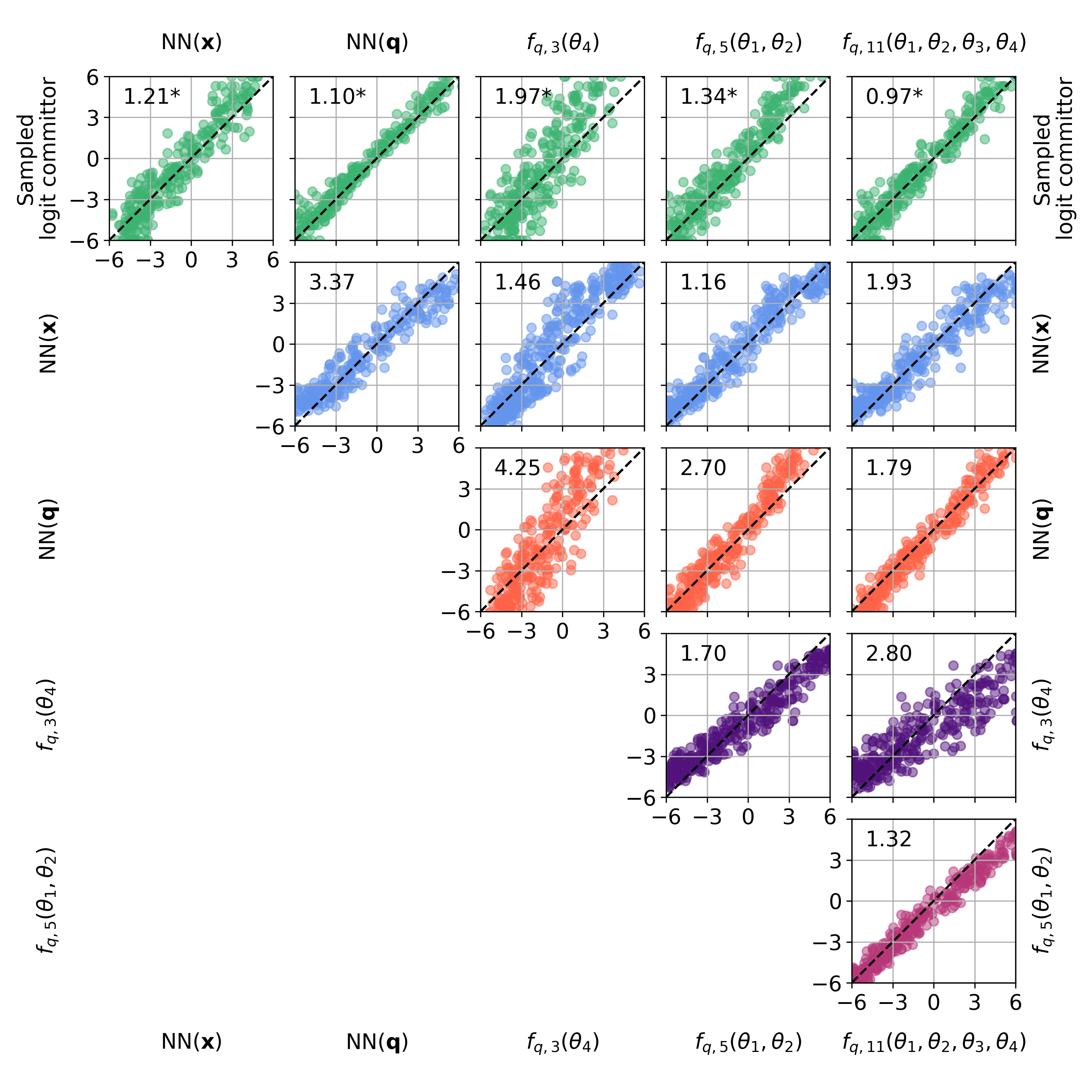}
    \caption{
        Pairwise correlations between the sampled logit committor values and our 5 logit committor models:
        NN($\mathbf{x}$): NN logit committor model based on atom pair distances,
        NN($\mathbf{q}$): NN logit committor model based on internal coordinates,
        and the SR solutions $f_{q,3}(\theta_4)$, $f_{q,5}(\theta_1, \theta_2)$, $f_{q,11}(\theta_1, \theta_2, \theta_3, \theta_4)$ for the logit committor prediction.
        The top-left value denotes the RMSE between the two logit committors.
        Values marked * do not include sampled committor values of zero (175 out of 438), for which the logit function is not defined.
        }
    \label{fig:logit_committor_comparison}
\end{figure}

\clearpage

\end{suppinfo}

%
%

\bibliography{literature}

\end{document}